\documentclass[a4paper,11pt]{amsart}
\usepackage[T1]{fontenc}

\usepackage[english]{babel}
\usepackage[utf8]{inputenc}
\usepackage{amsmath}

\newtheorem{theorem}{Theorem}

\theoremstyle{remark}
\newtheorem{remark}[theorem]{Remark}

\title{A formula for constructing Mignotte sequences}

\author{Marek Putresza}
\address{Faculty of Mathematics and Computer Science, Adam Mickiewicz University, Poznań, Poland}
\date{}

\begin{document}

\begin{abstract} 
	\noindent We present a new, direct and simple formula for constructing Mignotte sequences. \end{abstract}
	\maketitle
One of the secret sharing schemes is the Mignotte threshold secret sharing scheme, which uses the Mignotte sequences, defined as follows:

A $(k,n)$-Mignotte sequence, where $1<k<n$ are integers, is an increasing sequence 
\begin{equation*}
m_1 < \ldots < m_n
\end{equation*}
of pairwise relatively prime positive integers such that 
\begin{equation}
    \label{prodcondition}\prod_{i=1}^{k}m_{i}>\prod_{j=n-k+2}^{n}m_{j},
\end{equation}
i.e., the product of the smallest $k$ terms  is greater than the product of the $k-1$ largest ones.To ease the notation, put 
\begin{equation*}
 M=\prod_{i=1}^{k}m_{i}, \, N=\prod_{j=n-k+2}^{n}m_{j}.
 \end{equation*} 
Additionally, we require the quantity $\frac{M-N}{N}$ to be big.
 
The known method of constructing Mignotte sequences (see, e.g., \cite{K}, page 9) is indirect; it relies on finding special intervals with sufficiently many prime numbers. In this note, we prove the following theorem, which provides us with a direct and simple formula.   
 

\begin{theorem}\label{main}
 Let $q_1,\ldots,q_n$, $n\geq3$, be an increasing sequence of pairwise relatively prime positive integers. Define: 
\begin{equation}
    \label{defofP}P=\prod_{1\leq i<j\leq n}(q_j - q_i)
\end{equation}
Then
\begin{equation}
    \label{defofnewseq}P+q_1,\ldots, P+q_n
\end{equation}
is a $(k,n)$-Mignotte sequence for any $1<k<n$.
\end{theorem}


\begin{proof}[Proof]
First we check that (\ref{defofnewseq}) are pairwise relatively prime numbers. Compute for any $i<j$
\begin{align*}
    \gcd(P+q_i,P+q_j)
    =\gcd(P+q_i,P+q_j-(P+q_i))=\\
    =\gcd(P+q_i,q_j-q_i)
    \stackrel{\star}{=}\gcd(q_i, q_j-q_i)
    =\gcd(q_i, q_j)=1\\
\end{align*}  
where the equation marked $\stackrel{\star}{=}$ is valid since $q_j-q_i$ divides $P$. 

Now we check that the condition (\ref{prodcondition}) is satisfied, i.e.,
\begin{equation}
    \label{newprodproof}\prod_{i=1}^{k}(P+q_{i})>\prod_{j=n-k+2}^{n}(P+q_{j}).
\end{equation}
We analyze the case for $n=3, k=2$ separately. We will show that a stronger inequality
\begin{equation}\label{n3 stronger}
    (P+q_1)^2>P+q_3
\end{equation}\\
is satisfied.
Rewrite it as
\begin{equation}
    \label{ineqfor3}P^2+P(2q_1-1)+{q_1}^2-q_3>0.
\end{equation}
Define the polynomial
\begin{equation*}
	f(x)=x^2+x(2q_1-1)+{q_1}^2-q_3.
\end{equation*}
We will show that inequality \eqref{ineqfor3} is valid by proving that $P$ is greater than any root of $f$. Compute the discriminant of $f$,
\begin{equation*}
    \Delta_f=(2q_1-1)^2-4({q_1}^2-q_3)=4q_3-4q_1+1>0.
\end{equation*}
The larger root of $f$ equals
\begin{equation*}
    \frac{\sqrt{\Delta_f}-(2q_1-1)}{2}.
\end{equation*}
We will show that
\begin{equation}\label{n3 root}
    q_3-q_1>\frac{\sqrt{\Delta_f}-(2q_1-1)}{2}.
\end{equation}
Indeed, transforming \eqref{n3 root}, we get the following equivalent inequalities
\begin{equation*}
    2(q_3-q_1)+(2q_1-1)> \sqrt{\Delta_f}
\end{equation*}
\begin{equation*}
    2q_3-1> \sqrt{\Delta_f}
\end{equation*}
\begin{equation*}
    (2q_3-1)^2>\Delta_f
\end{equation*}
\begin{equation*}
    4(q_3)^2-4q_3+1>4q_3-4q_1+1
\end{equation*}
\begin{equation*}
    q_3(q_3-1)>q_3-q_1
\end{equation*}
The last inequality is valid since $q_3>q_1\geq1$. Thus \eqref{n3 root} holds, and we are done, since $ P\geq q_3-q_1$.\\

Now we prove \eqref{newprodproof} in general case  $n\geq4$. 
Factoring out $P$, and then dividing by $P^{k-1}$, we get the following equivalent inequalities
\begin{equation*}
    \prod_{i=1}^{k}P(1+\frac{q_{i}}{P})>\prod_{j=n-k+2}^{n}P(1+\frac{q_{j}}{P})  
\end{equation*}
\begin{equation*}
    P^k\prod_{i=1}^{k}(1+\frac{q_{i}}{P})>P^{k-1}\prod_{j=n-k+2}^{n}(1+\frac{q_{j}}{P})
\end{equation*}
\begin{equation}\label{n4 equiv}
    P\prod_{i=1}^{k}(1+\frac{q_{i}}{P})>\prod_{j=n-k+2}^{n}(1+\frac{q_{j}}{P}).
\end{equation}
To prove \eqref{n4 equiv}, we will show 
that a stronger inequality
\begin{equation}\label{n4 stronger}
P^{\frac{1}{k}}(1+\frac{q_{1}}{P})>(1+\frac{q_{n}}{P})
\end{equation}\\
is satisfied. Transforming it, we get
\begin{equation*}
    P^{\frac{1}{k}}(1+\frac{q_{1}}{P})-(1+\frac{q_{1}}{P})>(1+\frac{q_{n}}{P})-(1+\frac{q_{1}}{P})
\end{equation*}
\begin{equation*}
    (P^{\frac{1}{k}}-1)(1+\frac{q_{1}}{P})>\frac{q_{n}-q_{1}}{P}
\end{equation*}
\begin{equation}\label{proofendform}
    P(P^{\frac{1}{k}}-1)(1+\frac{q_{1}}{P})>q_{n}-q_{1}.
\end{equation}
We have \label{w1}$P\geq q_n-q_1$ by  (\ref{defofP}), and  obviously \label{w2}$(1+\frac{q_{1}}{P})>1$, since both $q_1$ and $P$ are positive. Thus, to prove \eqref{proofendform}, it remains to show that $P^{\frac{1}{k}}-1 > 1$, i.e., that \label{w3}$P^{\frac{1}{k}}>2$. Since the sequence $q_1,\ldots,q_n$ is increasing,  we get for any $n\geq 4$
\begin{equation*}
    P=\prod_{1\leq i<j\leq n}(q_j - q_i)\geq \prod_{1\leq i<j\leq n}(j-i)=\prod_{i=1}^{n-1}i!>2^{n-1}.
\end{equation*}
Hence indeed
\begin{equation*}
    P^{\frac{1}{k}}>2.
\end{equation*}
\end{proof}

\begin{remark}\label{uwaga}
Note that the procedure given by Theorem \ref{main} can be iterated as many times as one wishes. If the procedure is applied $t$ times, the resulting numbers are  
\begin{equation*}
    tP+q_1,\ldots, tP+q_n,
\end{equation*}
which asserts that the quantity $\frac{M-N}{N}$ can be as big as needed. Indeed, for $t$ large enough we get 
\begin{equation*}
    tP+q_1 \approx \ldots  \approx tP+q_n,
\end{equation*}
so 
\begin{equation*}
    \frac{M-N}{N} \approx tP+q_1 -1. 
\end{equation*}
The $P$ itself is large, since
\begin{equation*}
    \label{barnesg}P  > \prod_{i=1}^{n-1}i!>e^{n}
\end{equation*}
for $n>4$, where $e$ is the base of the natural logarithm. 

\begin{remark} Constructing an initial sequence of pairwise relatively prime positive integers, which is required in Theorem  \ref{main}, is very easy and can be done in many ways; consider, for example, the following elementary construction.\\	
	Let $q_{1}$ be a positive integer. Then the terms of the sequence defined recursively by 
	\begin{equation*}
		q_{k+1}=1+\prod_{i=1}^{k}q_{i}
	\end{equation*}
	are pairwise relatively prime.
\end{remark}

\end{remark}

\section*{Acknowledgements}
The author acknowledges the support from BESTStudentGRANT 119/39/UAM/0010. Words of gratefulness go to Stefan Bara\'{n}czuk for proposing Theorem \ref{main}  and the supervision. Special thanks go also to Tomasz Kościuszko, as an inspirer of our participation in the BESTStudentGRANT program. Lastly, we acknowledge Aleksander Tytus for discussions over the cryptographic side of this project, as well as suggestions that lead to simplification of the proof.

\bibliographystyle{plain}

\end{document}